\documentclass[twocolumn,showpacs,preprintnumbers,amsmath,amssymb]{revtex4}
\input psfig.sty

\usepackage{graphicx}
\usepackage{dcolumn}
\usepackage{bm}

\begin{document}

\def\be{\begin{equation}}
\def\ee{\end{equation}}
\def\a{\alpha}
\def\fr{\frac}
\def\berr{\begin{eqnarray}}
\def\err{\end{eqnarray}}

\def\apj{{\it Astrophys.~J.}}
\def\aj{{\it Astronom.~J.}}
\def\apjl{{\it Astrophys.~J.~Lett.}}
\def\prd{{\it Phys.~Rev.~D.}}
\def\prl{{\it Phys.~Rev.~Lett.}}
\def\plb{{\it Phys.~Letts.~B}}
\def\mnras{{\it Mon.~Not. R.~Astr.~Soc.}}
\def\ijmp{{\it Int.~J.~ Mod. Phys.}}
\def\nature{{\it Nature}}
\def\science{{\it Science}}
\def\AnA{{\it Astr. Astrophys.}}
\def\grg{{\it Gen. Rel. Grav.}}

\renewcommand{\a}{\alpha}
\renewcommand{\b}{\beta}
\newcommand{\g}{\gamma}
\newcommand{\G}{\Gamma}
\renewcommand{\d}{\delta}
\newcommand{\D}{\Delta}
\newcommand{\e}{\epsilon}
\renewcommand{\l}{\lambda}
\renewcommand{\L}{\Lambda}
\newcommand{\n}{\nabla}
\renewcommand{\o}{\omega}
\renewcommand{\O}{\Omega}
\renewcommand{\r}{\rho}
\newcommand{\s}{\sigma}
\renewcommand{\S}{\Sigma}
\renewcommand{\t}{\theta}
\renewcommand{\v}{\varphi}
\renewcommand{\P}{\Phi}
\newcommand{\del}{\partial}

\title{The angular size - redshift relation in power-law cosmologies}

\author{Deepak Jain}\email{deepak@ducos.ernet.in}
\affiliation{Deen Dayal Upadhyaya College, University of Delhi, Delhi - 110015, India}

\author{Abha Dev} \email{abha@ducos.ernet.in}
\affiliation{Department of Physics and Astrophysics, University of Delhi,
Delhi - 110007, India}

\author{J. S. Alcaniz} \email{alcaniz@astro.washington.edu}
\affiliation{Astronomy Department, University of Washington, Seattle,
Washington, 98195-1580, USA}

\date{\today}

\begin{abstract}

A linear evolution of the cosmological scale factor is a feature in several models designed to solve the
cosmological constant problem via a coupling between scalar or tensor classical fields to the space-time
curvature as well as in some alternative gravity theories. In this paper, by assuming a general time
dependence of the scale factor, $R \sim t^{\alpha}$, we investigate observational constraints on the
dimensionless parameter $\alpha$ from measurements of the angular size for a large sample of milliarcsecond
compact radio sources. In particular, we find that a strictly linear evolution, i.e., $\alpha \simeq 1$ is
favoured by these data, which is also in agreement with limits obtained from other independent cosmological
tests. The dependence of the critical redshift $z_m$ (at which a given angular size takes its minimal value)
with the index $\alpha$ is briefly discussed.

\end{abstract}

\maketitle

\section{Introduction}

Over the past few years, an impressive convergence of observational facts have led cosmologists to search for
alternative cosmologies. Among these facts, the most ramarkable finding surely comes from distance
measurements of
type Ia supernovae (SNe Ia) that suggest that the expansion of the universe is speeding up, not slowing down
\cite{perlmutter}. Such a result, when combined with the latest Cosmic Microwave Background (CMB) data and
clustering estimates, seems to provide a compelling evidence for a non-zero cosmological constant $\Lambda$
\cite{efs}.

On the other hand, it is also well known that the same welcome properties that make models with a relic
cosmological constant ($\Lambda$CDM) our best description of the observed universe also result in a serious
fine
tuning problem \cite{wein}. The basic reason is the widespread belief that the early universe evolved through
a
cascade of phase transitions, thereby yielding a vacuum energy density which is presently 120 orders of
magnitude
smaller than its value at the Planck time. Such a discrepancy between theoretical expectations and empirical
observations contitutes a fundamental problem at the interface uniting astrophysics, cosmology and particle
physics. In the last years, several attempts have been done in order to alleviate the cosmological constant
problem. For example, in the so-called dynamical $\Lambda(t)$ scenarios (or deflationary cosmology), the
cosmological term is a function of time and its presently observed value is a remnant of the primordial
inflationary/deflationary stage \cite{ozer}. Other examples are scenarios in which the evolution of classical
fields are coupled
to the curvature of the space-time background in such a way that their contribution to the energy density
self-adjusts to cancel the vacuum energy \cite{dolgov}, as well as some recent ideas of a SU(2) cosmological
instanton dominated universe \cite{allen}. At least in the two latter exemples, an interesting feature is a
power-law growth for the cosmological scale factor $R(t) \sim t^{\alpha}$, where $\alpha$ 
is determined by observational data and takes values within the interval $0 \leq \alpha \leq \infty$. A
linear
evolution of the scale factor is also supported in some alternative gravity theories, e.g., non-minimally
coupled
scalar-tensorial theories \cite{linear}, as well as in standard model with a specially chosen equation of
state
\cite{kolb}.

The motivation for seriously investigating these power-law scenarios comes from several considerations. For
example, for $\alpha \geq 1$ such models do not suffer the horizon problem. Moreover, the scale factor in such
theories does not constrain the matter
density
parameter and, therefore, they are free of the flatness problem. There are also observational motivations for
considering power-law cosmologies. For $\alpha \geq 1$, the predicted age of the universe is $t_o \geq
H_o^{-1}$,
i.e., at least $30-50\%$ greater than the prediction of the flat standard model (without cosmological
constant),
thereby making such a universe confortably in agreement with the recent age estimates of globular clusters and
high-$z$ redshift galaxies \cite{dev1}. Recently, it was shown that such models are also compatible with the
currently data of SNe Ia for a power index $\alpha \simeq 1$ \cite{dev} (see, however, \cite{steigman} for a
discussion involving SNe Ia and primordial nucleosynthesis constraints).

In this paper we explore the prospects for constraining the power-law index $\alpha$ from the angular size
measurements of high-$z$ milliarcsecond radio sources. We also study the influence of this dimensionless
parameter
on the minimal redshift at which the angular size of an extragalactic source takes its minimal value. For the
sake of simplicity and also motivated by the latest results of CMB analyses we focus our attention on flat
scenarios.
We show
that a good agreement between theory and
observation
is possible if $\alpha = 1.0 \pm 0.3$ at $68\%$ c.l. with the characteristic lenght of the sources of the
order
of $l \simeq 26$ pc (for $H_o =  72 \rm{km} \rm{s^{-1}}
\rm{Mpc^{-1}}$).

This paper is organized as follows. In section II some basic assumptions and distance formulas are presented.
The
dependence of the minimal redshift $z_m$ on the index $\alpha$ is studied in section III. In section IV we
analyse
the constraints from angular size data on this class of cosmologies and compare them with other independent
limits.
In section V our main conclusions are presented.

\section{Power-law cosmologies: basic equations}

Let us now consider the flat Friedmann-Robertson-Walker (FRW) line element 
\begin{equation}
ds^2 = c^{2}dt^2 - R^{2}(t) \left[{d\xi^{2}} + \xi^{2} (d
\theta^2 + \rm{sin}^{2} \theta d \phi^{2})\right],
\end{equation}
where $\xi$, $\theta$, and $\phi$ are dimensionless comoving coordinates and $R(t)$ is the cosmological scale
factor.

We study a general class of power law cosmology in which the scale factor is given in terms of the arbitrary
dimensionless parameter $\a$
\begin{equation}
R(t) \propto \fr{c}{H_0}\left(\fr{t}{t_0}\right)^{\a} .
\label{eq:ansatz}
\end{equation}

The expansion rate of the universe is described by the Hubble parameter, $H(t) =
\dot{a}/a ={\a}/{t}$ while the {\it present} expansion rate is
defined by a Hubble {\it constant}, equal in this model to
\begin{equation}
H_o=\a/t_0
\end{equation}
(here and subsequently the subscript ``o" refers to its present value). By comparing the above expression with
some
recent estimates of the age parameter, it is possible to obtain directly limits on the dimensionless parameter
$\alpha$. For example, by assuming $t_o = 13 \pm 2$ Gyr as a median value for the age estimates of globular
clusters and using $H_o = 72 \pm 8 \rm{km} \rm{s^{-1}} \rm{Mpc^{-1}}$, in accordance with the final results of
the
\emph{Hubble Space Telescope} Key Project \cite{freedman}, we find $\alpha = 0.96 \pm 0.19$.

As usual, the scale factor and the redshift are related by $a/a_o = (1 + z)^{-1}$ while the dimensionless
Hubble
parameter now takes the form
\begin{equation}
h(z) \equiv \frac{H(z)}{H_0}=(1+z)^{1/\a}.
\label{eq:dimensionless_hubble}
\end{equation}

From the above equations, it is straightforward to show that the comoving distance $\xi(z)$ for flat
geometries is given by
\begin{equation}
\xi(z) =  \frac{c \alpha}{R_o H_o (1 - \alpha)} \left[1 - (1+z)^{\frac{\alpha -1}{\alpha}}\right],
\end{equation}
which, in the limit $\a \rightarrow 1$, reduces to
\begin{equation}
\xi(z) = \frac{c}{R_oH_o}\rm{ln}(1 + z).
\end{equation}

The angular size redshift relation for a rod of intrinsic lenght $l$ is easily obtained by integrating the
spatial
part of Eq. (1) for $\xi$ and $\phi$ fixed \cite{sand}. One finds
\begin{eqnarray}
\theta(z) = {l (1 + z) \over  R_o\xi(z)} =
 {D(1 + z)(1 - \alpha) \over \alpha\left[1 - (1+z)^{\frac{\alpha -1}{\alpha}}\right]},
\end{eqnarray}
where the characteristic lenght $l$ is measured in parsecs (for compact radio sources) and the characteristic
angular scale $D = lH_o/c$ is
given in milliarcsecond (mas).

\section{The critical redshift}

As is well known, the existence of a critical redshift $z_m$ on the angular size - redshift relation may be 
qualitatively understood in terms of the cosmological expansion. The light observed today from a source at
a given redshift $z$ was emmited when the object was closer (for a detailed discussion see \cite{janis}).
Although nearby objects are not affected, a fixed angular size of extragalactic sources at high-$z$ is seen
iniatially
decreasing to a minimal value, say, $z_m$, and afterwards increasing to higher redshifts.
Although this minimal redshift test cannot discriminate by itself among different cosmological models
(different
scenarios may predict the same $z_m$ values) \cite{alcaniz}, a precise determination of $z_m$ or,
equivalently, the
corresponding minimal angular size value $\theta(z_m)$, may constitute, when combined with other cosmological
tests, a powerful tool to check the validity of realistic world models. Such an effect was first predicted by
Hoyle
\cite{hoyle}, originally aiming at distinguishing the steady state and Einstein-de Sitter cosmologies. Later
on,
the accumulated evidence against the original version of the steady state model have put such a scenario
aside, and more
recently
the same is occurring with the theoretically favoured critical density FRW model \cite{teg}.

\begin{figure}
\centerline{\psfig{figure=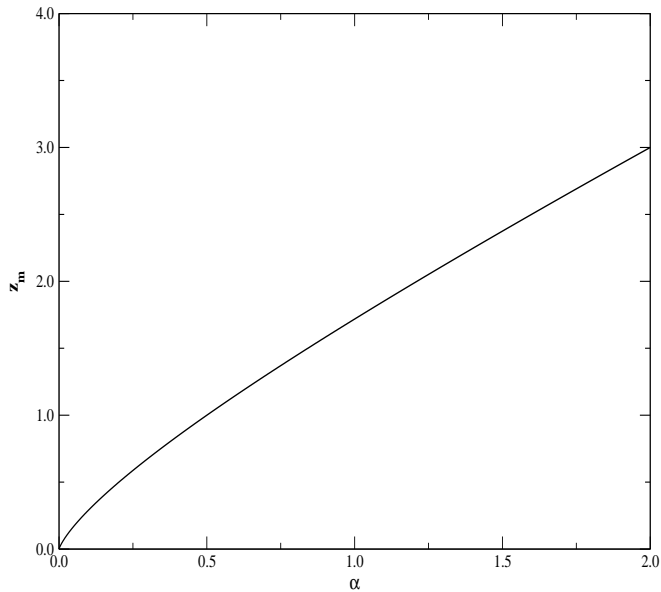,width=3.5truein,height=3.3truein,angle=-90}
\hskip 0.1in}
\caption{Critical redshift $z_m$ as a function of the dimensionless parameter $\alpha$. As expected, for
$\alpha = 2/3$ the standard prediction, $z_m = 5/4$, is recovered. For $\alpha \rightarrow 1$, the critical
redshift approaches 1.72.}
\end{figure}

The redshift $z_m$ at which the angular size takes its minimal value is the one cancelling out the derivative
of
$\theta$ with respect to $z$. From Eq. (7), we find
\begin{equation}
z_m = \alpha^{\frac{\alpha}{\alpha - 1}} - 1,
\end{equation}
which provides $z_m = e -1 \simeq 1.72$ in the limit $\alpha \rightarrow 1$. Such a value is similar to the
one
predicted by an open FRW model with the matter density parameter of the order of $\Omega_m \sim 0.4$ (see
Table 1
of \cite{alcaniz}) as well as to the one predicted by a flat model with cosmological constant and $\Omega_m
\simeq 0.2$ (see Table 1 of \cite{krauss}). In Fig. 1 we show the diagram $z_m$ as a function of the
dimensionless parameter $\alpha$. Clearly, from the above equation, $z_m$ is an increasing function of
$\alpha$.
Note still that for values of $\alpha \lesssim 0.5$ the minimal redshift takes small values ($z_m \lesssim
1$),
which is not observed from the current data. As expected, for $\alpha = 2/3$ the standard prediction, $z_m =
5/4$,
is readily recovered.

\section{Constraints from high-$z$ angular size measurements}

In this section we study the constraints on the parameter $\alpha$ from the angular size measurements of
high-$z$
milliarcsecond radio sources. To place such constraints we use the $\theta (z)$ data compiled by Gurvits {\it
et
al.} \cite{gurv}. This data set, originally composed by 330 sources distributed over a wide range of redshift
($0.011 \leq z \leq 4.72$), was reduced to 145 sources with spectral index between $[-0.38,0.18]$ and
total
luminosity $Lh^{2} \geq 10^{26}$ W/Hz in order to minimize any possible dependence of the angular size on the
spectral index and/or linear size on luminosity  \cite{gurv}. This new subsample was distributed into 12 bins
with 12-13
sources
per bin. In our analysis we assume that possible evolutionary effects can be removed from this sample since
compact
radio jets are (i) typically less than 100 pc in extent, and therefore, their morphology and kinematics do not
depend considerably on the intergalactic medium and (ii) they have ages of the order of years, i.e., much
smaller
than the cosmological scale of time $H_o^{-1}$ \cite{kell}. This particular data set has been extensively used
in the
recent literature, with several authors aiming mainly at constraining different quintessence scenarios
\cite{myself}.

\begin{figure}
\centerline{\psfig{figure=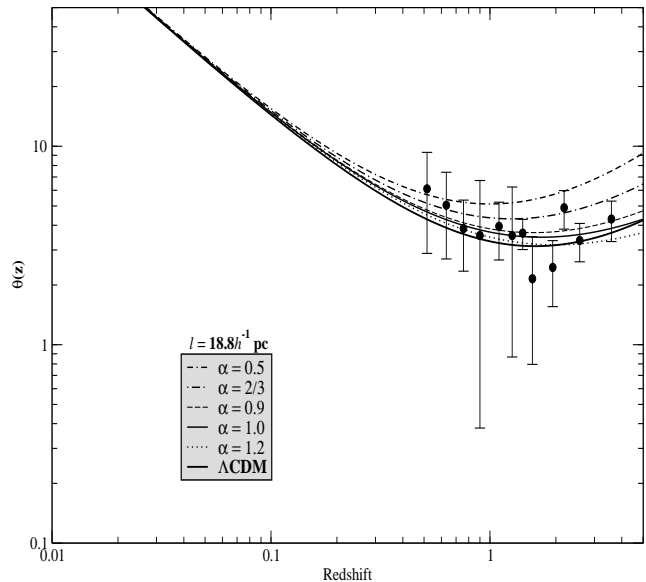,width=3.5truein,height=3.3truein,angle=-90}
\hskip 0.1in}
\caption{Angular size versus redshift for selected values of $\alpha$. The data points correspond to 145
sources binned into 12 bins \cite{gurv}. In all curves the characteristic angular size has been fixed at $l =
18.8h^{-1}$ pc ($D = 1.28$ mas). The thick line is the $\Lambda$CDM prediction.}
\end{figure}

To determine the confidence regions in the plane $D - \alpha$ we use a $\chi^{2}$ minimization for the range
of
$\alpha$
and $D$ spanning the interval [0,1] and [0.01,2], 
respectively
\begin{equation}
\chi^{2} = \sum_{i=1}^{12}{\frac{[\theta(z_i, D, \alpha) - \theta_{oi}]^{2}}{\sigma_i^{2}}},
\end{equation}
where $\theta(z_i, D, \alpha)$ is given by Eq. (7) and $\theta_{oi}$ stands for the observed values of the
angular
size with errors $\sigma_i$ of the $i^{th}$ bin in the sample.

Figure 2 shows a log-log plot of the angular size versus redshift for power-law cosmologies with some selected
values of the dimensionless parameter $\alpha$. For the sake of comparison, the current favoured cosmological
model, namely, a flat scenario with $\sim 70\%$ of the critical energy density dominated by a cosmological
constant
($\Lambda$CDM) is also shown. In Fig. 3 we show the contours of constant likelihood (68\% and 95\%) in the
plane
$\alpha - D$. Note that the data set allows a large interval for the parameter $\alpha$, which shows the
impossibility
of placing very restrictive constraints on the index $\alpha$ from the current $\theta(z)$ data. From this
analysis we obtain $\alpha = 1.006$ and $D \simeq 1.28$ ($l \simeq 18.8h^{-1}$ pc) as the best fit values.
This particular value
of $\alpha$ is very close to the one obtained in Sec. II by using recent age estimates of
globular clusters and also supports the idea of a strictly linear evolution of $R(t)$ (i.e., $R(t) \propto
t$). In Fig. 4, by
marginalising over the characteristic angular scale $D$, we show the normalised likelihood
curve (${\cal{L}}/{\cal{L}}_{max}$) as a
function of the parameter $\alpha$. The best-fit value for this parameter corresponds to the maximum value of
the likelihood function
\begin{equation}
{\cal{L}}(\a) = \int \exp(-\chi^2(\a,D))\,dD.
\end{equation}
As can be seen from that figure, the likelihood is a sharply function of
$\alpha$ with its maximum value centered at $\alpha = 1.006$.
We also obtain  $0.704 \le \alpha \le 1.312$ within $68\%$ c.l. and 
$0.428 \le \alpha \le 1.908$ within $95\%$ c.l..

\begin{figure}
\centerline{\psfig{figure=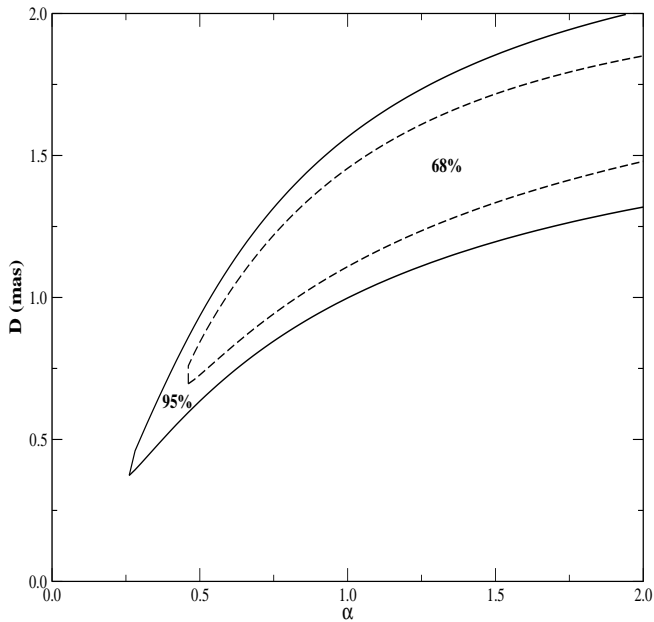,width=3.5truein,height=3.5truein,angle=-90}
\hskip 0.1in}
\caption{Confidence regions in the $\alpha - D$ plane according to the updated sample of angular size data of
Gurvits {\it et al.} \cite{gurv}. The contours correspond to $68\%$ (dashed) and $95\%$ (solid) confidence
levels.}
\end{figure}

At this point, it is also interesting to compare our results with some other independent constraints on the
power-law index $\alpha$. For example, Dev {\it et al.} \cite{dev1} used SNe Ia data to show that a good
agreement between these data and
the class of power-law cosmologies studied here is possible for values of $\alpha = 1.0 \pm 0.04$. For a open
scenario a study of the statistical properties of gravitational lenses
provided $\alpha = 1.13^{+0.4}_{-0.3}$ at 68\%
c.l. while the predicted number of lensed quasars for the same sample
implied
$\alpha = 1.09 \pm 0.3$ \cite{dev}. Age estimates of high-$z$ galaxies require a lower limit of
$\alpha \geq 0.8$
\cite{dev}. All these determinations agree satisfactorily with the limits obtained from the analysis presented
in this paper.

\begin{figure}
\centerline{\psfig{figure=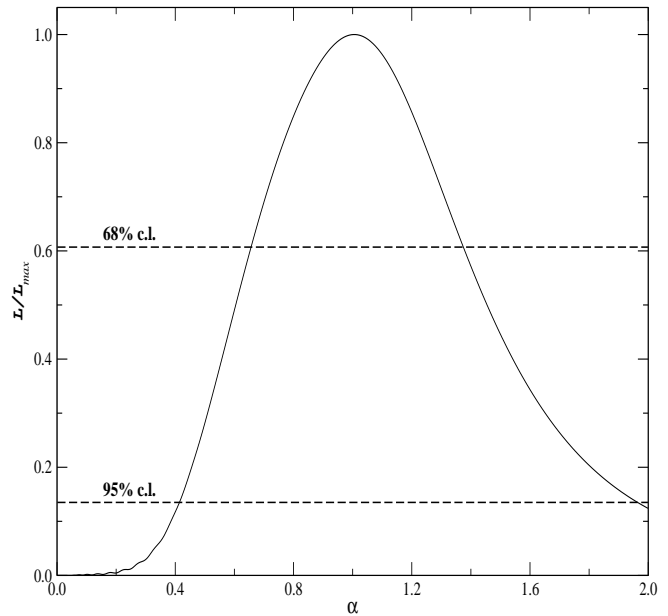,width=3.5truein,height=3.5truein,angle=-90}
\hskip 0.1in}
\caption{The likelihood curve, normalised with the maximum likelihood value, is shown as a function of the
index $\alpha$. The remaining parameter $D$ (or equivalently, $l$) has been marginalised. As indicated, the
horizontal dashed lines correspond to confidence limits of 68$\%$ and $95\%$.}
\end{figure}

\section{conclusion}

We have studied some properties of the angular size - redshift relation in a general class of power-law
cosmologies in which the scale factor is expressed as a function of an arbitrary dimensionless parameter
$\alpha$. We
have 
investigated the influence of this parameter on the redshift $z_m$ at which the angular size takes its minimal
value. Moreover, by using measurements of the angular size for a large sample of milliarcsecond compact radio
sources we also have placed new constraints on the power-law index $\alpha$. In particular, after
marginalising over the
characteristic angular scale $D$ we have found $\alpha = 1 \pm 0.3$
at 68\% c.l., a result that agrees very well with other independent determinations based on different methods.
We emphasize that if a definitive agreement between the contraints derived from classical cosmological tests
and
primordial nucleosythesis could be shown, this class of power-law cosmologies would constitute an interesting
alternative to the standard cosmology.

\begin{acknowledgments}
The authors are grateful to Professor L. I. Gurvits for sending his compilation of the angular size data and
to R. G. Vishwakarma and E. Bentivegna for useful discussions. JSA
is 
supported by the Conselho Nacional de Desenvolvimento Cient\'{\i}fico e Tecnol\'{o}gico (CNPq - Brasil) and
CNPq
(62.0053/01-1-PADCT III/Milenio).
\end{acknowledgments}


\begin{thebibliography}{30}
\bibitem{perlmutter} S. Perlmutter {\it et al.}, Nature, {\bf{ 391}}, 51 (1998); S. Perlmutter {\it
et al.}, Astrophys. J. {\bf{517}}, 565 (1999); A. Riess {\it et al.}, Astron. J. {\bf{116}}, 1009 (1998)
\bibitem{efs} G. Efstathiou {\it et al.}, astro-ph/0109152
\bibitem{wein} S. Weinberg, Rev. Mod. Phys. {\bf{61}}, 1 (1989);
V. Sahni and A. A. Starobinsky, Int. J. Mod. Phys. {\bf{D9}}, 373 (2000)
\bibitem{ozer}  M. Ozer and  M. O. Taha Phys. Lett. {\bf{B171}}, 363 (1986); O. Bertolami, Nuovo Cimento,
{\bf{B93}}, 36 (1986); K. Freese {\it et al.}, Nucl. Phys.
{\bf{B287}}, 797 (1987); W. Chen and Y.-S. Wu, Phys. Rev. {\bf{D41}}, 695 (1990); J. C. Carvalho,  J. A. S.
Lima and I. Waga, Phys. Rev. {\bf{D46}} 2404 (1992); J. M. Salim and I. Waga, Class. Quant, Grav. {\bf{10}},
1767 (1993); J. Matyjasek, Phys. Rev. {\bf{D51}}, 4154 (1995); O. Bertolami and P. J. Martins, Phys. Rev.
{\bf{D61}}, 064007 (2000); R. G. Vishwakarma, Gen. Rel. Grav. {\bf{33}}, 1973 (2001); J. S. Alcaniz and J.M.F.
Maia, Phys. Rev. D (in press) astro-ph/0212510; J. V. Cunha, J. A. S. Lima and J. S. Alcaniz, Phys. Rev.
{\bf{D66}}, 023520 (2002). astro-ph/0202260  
\bibitem{dolgov} A. D. Dolgov in {\em The Very Early Universe}, ed. 
G. Gibbons, S. W. Hawking, and S. T. Tiklos (Cambridge University Press, 
1982); F. Wilczek, Phys. Rep. {\bf 104} (1984) 143; R. D. Peccei, J. Sola, and C. Wetterich, 
Phys. Lett. {\bf B195} (1987) 18; L. H. Ford, Phys. Rev. {\bf D35} (1987) 
2339; S. M. Barr and D. Hochberg, Phys. Lett. {\bf B211} (1988) 49;
Y. Fujii and T. Nishioka, Phys. Rev. {\bf D42} (1990) 361; Phys.Lett. 
{\bf B254} (1991) 347; A. D. Dolgov,  Phys. Rev. {\bf D55} (1997) 5881.
\bibitem{allen} R. E. Allen, astro-ph/9902042
\bibitem{linear} D. Lohiya and M. Sethi, Class. Quant. Grav. {\bf{16}}, 1545 (1999)
\bibitem{kolb} E. W. Kolb, \apj {\bf{344}}, 543 (1989)
\bibitem{dev1} A. Dev, M. Safonova, D. Jain and D. Lohiya, Phys. Lett. {\bf{B548}}, 12 (2002)
\bibitem{dev} A. Dev, M. Sethi and D. Lohiya, Phys. Lett. {\bf{B504}}, 207 (2001)
\bibitem{steigman} M. Kaplinghat, G. Steigman, I. Tkachev and T. Walker, Phys. Rev. {\bf{D59}}, 043514 (1999);
M. Sethi, A. Batra and D. Lohiya, Phys. Rev. {\bf{D60}}, 108301 (1999); M. Kaplinghat, G. Steigman and T.
Walker, Phys. Rev. {\bf{D61}}, 103507 (2000)
\bibitem{freedman} W. L. Freedman {et al.}, \apj {\bf{553}}, 47 (2001)
\bibitem{sand} A. R. Sandage, ARA\&A, {\bf{26}}, 56 (1988)
\bibitem{janis} A. I. Janis, Amer. Journ. Phys., {\bf{54}} 1008 (1986)
\bibitem{alcaniz} J. A. S. Lima and J. S. Alcaniz, Astron. Astrop. {\bf{357}}, 393 (2000); Gen. Rel. Grav.
{\bf{32}}, 1851 (2000)
\bibitem{hoyle} F. Hoyle, in {\it Radio Astronomy. IAU Symp. No. 9}, Ed. R. N. Bracewell, Stanford Univ.
Press., 529 (1959)
\bibitem{teg} M. Tegmark, Science {\bf{296}}, 1427 (2002), astro-ph/0207199; P. J. E. Peebles and B. Ratra,
astro-ph/0207347;  M. S. Turner, Int. J. Mod. Phys. {\bf{A17}}, 3446 (2002)
\bibitem{krauss} L. M. Krauss and D. N. Schrammm, \apj {\bf{405}}, L43
\bibitem{gurv} L. I. Gurvits, K. I. Kellermann and S. Frey, Astron. Astrop. {\bf{342}}, 378 (1999)
\bibitem{kell} K. I. Kellermann, Nature {\bf{361}}, 134 (1993)
\bibitem{myself} J. C. Jackson and M. Dodgson, MNRAS, {\bf{278}}, 603; MNRAS {\bf{285}}, 806, (1997); R. G.
Vishwakarma, Class. Quant. Grav. {\bf{18}}, 1159 (2001); J. A. S. Lima and J. S. Alcaniz, \apj {\bf{566}}, 15
(2002). astro-ph/0109047; J. S. Alcaniz, Phys. Rev. D {\bf{65}}, 123514 (2002). astro-ph/0202492; Z. Zhu and
M. Fujimoto, \apj
{\bf{581}}, 1 (2002); G. Chen and B. Ratra, \apj, {\bf{582}}, 586 (2003)

\end{thebibliography}

\end{document}